\documentclass[preprint2]{aastex62}

\received{2018 June 15}
\revised{2018 September 18}
\accepted{2018 October 15}

\submitjournal{ApJ}

\shorttitle{The evolution of the WA in IRAS\,17020+4544}
\shortauthors{Sanfrutos et al.}

\usepackage[utf8]{inputenc}
\usepackage{lipsum} % Used for inserting dummy 'Lorem ipsum' text into the template
\usepackage[acronym,nomain]{glossaries} % Para usar acrónimos. Muestra el nombre completo sólo la primera vez que se escribe.  
\newacronym{los}{LOS}{line of sight}
\newacronym{blr}{BLR}{broad line region}
\newacronym{nlr}{NLR}{narrow line region}
\newacronym{agn}{AGN}{active galactic nuclei}
\newacronym{dof}{dof}{degrees of freedom}
\newacronym{sed}{SED}{spectral energy distribution}
\newacronym{epic}{EPIC}{European Photon Imaging Camera}
\newacronym{ew}{EW}{equivalent width}
\newacronym{ufo}{UFO}{ultra-fast outflow}
\newacronym{rgs}{RGS}{Reflection Grating Spectrometer}
\newacronym{nls1}{NLS1}{narrow line Seyfert\,1}
\newacronym{gr}{GR}{general relativity}
\newacronym{smbh}{SMBH}{supermassive black hole}
\newacronym{isco}{ISCO}{innermost stable circular orbit}
\newacronym{fwhm}{FWHM}{full width at half maximum}
\newacronym{uv}{UV}{ultraviolet}
\newacronym{euv}{EUV}{extreme ultraviolet}
\newacronym{lcdm}{$\Lambda$CDM}{$\Lambda$ cold dark matter}
\newacronym{mos}{MOS}{metal--oxide--semiconductor}
\newacronym{heasarc}{HEASARC}{High Energy Astrophysics Science Archive Research Center}
\newacronym{heg}{HEG}{High Energy Gratings}
\newacronym{wa}{WA}{warm absorber}
\newacronym{snr}{SNR}{Supernova Remnants}
\newacronym{ism}{ISM}{Interstellar Medium}
\newacronym{letg}{LETGS}{Low Energy Transmission Grating Spectrometer}
\newacronym{asca}{ASCA}{Advanced Satellite for Cosmology and Astrophysics}
\newacronym{vlbi}{VLBI}{very long baseline interferometry}

\def\xmm{{\it XMM-Newton}}

\def\mnras{MNRAS}       % Monthly Notices of the Royal Astronomical Society 
\def\apj{ApJ}           % The Astrophysical Journal 
\def\apjs{ApJS}         % The Astrophysical Journal, Supplement 
\def\aap{A\&A}          % Astronomy & Astrophysics 
\def\aaps{A\&AS}        % Astronomy and Astrophysics Supplement Series
\def\apjl{ApJ Letters}  % The Astrophysical Journal Letters 
\def\araa{ARA\&A}       % Annual Review of Astronomy & Astrophysics 
\def\nat{Nature}        % Nature
\def\nar{New Astron. Rev.} % New Astronomy Reviews
 % ...

\newcommand{\xr}{X-ray}
\newcommand{\kms}{km\,s$^{-1}$}
\newcommand{\ie}{i.e.}
\newcommand{\eg}{e.g.}

\newcommand{\iras}{IRAS\,17020+4544}

\begin{document}

\title{THE EVOLUTION OF THE WARM ABSORBER REVEALS A SHOCKED OUTFLOW IN THE NARROW LINE SEYFERT 1 GALAXY IRAS\,17020+4544}

\correspondingauthor{Mario Sanfrutos}
\email{sanfrutoscm@astro.unam.mx}

\author[0000-0002-2616-544X]{Mario Sanfrutos}
\affiliation{Instituto de Astronomía, Universidad Nacional Autónoma de México, A. P. 70264, 04510 CDMX, México}

\author{Anna Lia Longinotti}
\affiliation{CONACYT--Instituto Nacional de Astrofísica, Óptica y Electrónica, Luis E. Erro 1, Tonantzintla, Puebla, C.P. 72840, México}

\author{Yair Krongold}
\affiliation{Instituto de Astronomía, Universidad Nacional Autónoma de México, A. P. 70264, 04510 CDMX, México}

\author{Matteo Guainazzi}
\affiliation{ESTEC/ESA, Keplerlaan 1, 2201AZ Nordwijk, The Netherlands}

\author[0000-0003-0543-3617]{Francesca Panessa}
\affiliation{INAF Istituto di Astrofisica e Planetologia Spaziali di Roma, via del Fosso del Cavaliere 100, 00133, Roma, Italy}

\begin{abstract}
We present the analysis of grating spectra of the Narrow Line Seyfert 1 Galaxy \iras\ observed by \xmm\ in 2004 and 2014. 
In a previous work on these data, we reported the discovery of a multi-component ultra-fast outflow that is capable of producing feedback in the host galaxy. 
We also reported the presence of a slow, multi-phase warm absorber. In this follow-up paper, we confirm that this low velocity absorber can be modeled by four layers of ionized gas. When crossing our line-of-sight, this gas presents peculiar changes along the 10-yr time scale elapsed between the two observations obtained by \xmm .
While two of such components are almost stationary, the other two are found inflowing and outflowing with significant variations in velocity and ionization between 2004 and 2014.
The luminosity and spectral shape of the central source remain practically unvaried. 
We propose that the presence of the fast wind and of the variable warm absorber can be interpreted in the framework of  a ``shocked outflow'', where the peculiar variability pattern of the low-velocity components might arise from instabilities in the shocked gas. 
\end{abstract}


\section{Introduction} \label{sec:intro}

The warm absorber (WA) was first proposed by \cite{1984ApJ...281...90H} as a cloud or shell of photoionized material in active galactic nuclei (AGN), in order to explain the \xr\ spectrum of the QSO MR\,2251\,--\,178, whose main signature was first interpreted to be the K-edge of O${\textsc{vii}}$ at 739\,eV. 

In early samples of Seyfert Galaxies observed with the CCD detectors on board the Advanced Satellite for Cosmology and Astrophysics (\textit{ASCA}), at least half of the type~1 AGN show high-ionization oxygen K-shell absorption edges (the already mentioned O${\textsc{vii}}$, and O${\textsc{viii}}$ at 0.87\,keV) typical of the WA \citep{1997MNRAS.286..513R, 1998ApJS..114...73G}; 
today we know that absorption  is produced by a realm of atomic transitions that result in a series of absorption lines from ionized elements \citep[see e.g.:][]{2003A&A...403..481B, 2000A&A...354L..83K, 2000ApJ...535L..17K, 2002ApJ...574..643K, 2003ApJ...597..832K, 2005ApJ...620..165K, 2005ApJ...622..842K}. 
Most features revealing the presence of WAs are imprinted on the soft \xr\ spectra of AGN. 
Ionized gas along the line of sight of Seyfert Galaxies has provided an excellent diagnostic for the properties and conditions of photo ionized material intrinsic to AGN. 
The origin of the WA has been sometimes related to an outflow arising from the dusty torus \citep{2005A&A...431..111B, 2001ApJ...561..684K, 2014MNRAS.437.1776M}. 
Results by \cite{2012A&A...539A.117K} on \xmm\ observations of Mrk 509 point to an origin of the WA components in the narrow line region (NLR) or torus region of this object.
In other cases, the response of the ionization state of the gas and, in general, variability of the absorber parameters have suggested that the inner accretion disk can be identified as the outflow launching region \citep{2014Sci...345...64K, 2007ApJ...659.1022K, 2013ApJ...766..104L}. Mixed situations where different components of the wind are launched in distinct regions of the AGN show that we are still far from reaching a homogenous picture to explain the absorber phenomenology. For instance, this was the case reported by  \cite{2016MNRAS.457..510S} in ESO\,323-G77, where we found one variable, ``disk-like'' WA component from the dust-free broad line region (BLR), and one persistent, ``torus-like'' component from the dusty, clumpy torus, showing absorption  variability at both short- and long-time scales. 

The velocity to which the ionized gas is accelerated is used to separate the standard, ``slow'' WAs with $v_{\rm out}$~$\le$ 10$^4$~\kms\ from the more recently discovered ultra-fast outflows (UFOs), which reach outflow velocities of 0.1-0.3~{\it c} \citep{2012MNRAS.422L...1T, 2013MNRAS.430...60G} and which seem to provide a key ingredient for triggering energy-driven outflows at large scale \citep{2015A&A...583A..99F, 2015Natur.519..436T}. 

There have been attempts to explore the relation between WAs and  UFOs \citep{2016MNRAS.457.3896L, 2013MNRAS.430.1102T}. Nevertheless, the paucity of sources where these two components of the inner \xr\ outflow are observed at the same time with comparable detail and signal to noise does not allow firm conclusions to be reached. 

\begin{sloppypar}
\iras\ is a narrow line Seyfert\,1 (NLS1) galaxy \citep{1996ApJS..106..341M, 1997A&A...320..395W} at redshift {\it z}=0.0604 \citep{1992A&AS...96..389D}, with a central supermassive black hole (SMBH) of mass $M_{\rm BH} \sim 5.9\times 10^{6} M_{\odot}$ \citep{2001A&A...377...52W}. Its \xr\ luminosity is $\sim 1.5\times 10^{44}$\,erg\,s$^{-1}$, from which a bolometric luminosity of $\sim5.2\times 10^{44}$\,erg\,s$^{-1}$ is estimated by assuming a correction {\it k}=10 \citep{2004MNRAS.351..169M}. 
\end{sloppypar}

The presence of a dusty WA in this galaxy was first suggested by \cite{leighly97} based on data obtained by the \textit{ASCA} satellite and later confirmed by \cite{1998A&A...331L..49K} using data from the \textit{ROSAT} mission. 
The first spectroscopic study of \iras\ with modern \xr\ observatories is reported by \cite[][from now on L15]{longinotti2015ApJ}, who discovered  a multi-component  UFO via detection of absorption lines in the \xmm\ grating spectrum.  
The ionized gas  was found to be outflowing  at sub-relativistic velocities ($\sim$\,23,000-34,000~\kms ) and to span low to moderate ionization ($\log{U}$ from $\sim$\,$-2$ to $\sim$\,$2.6$). The equivalent hydrogen column density of all the components  also spans a wide range of values  ($\log{N_{\rm H}} \sim$\,$20$-$24$\,cm$^{-2}$), with one of them being massive enough to 
possibly %MG
enable feedback from this wind onto the host galaxy.

The \xmm\ spectra also confirmed the presence of a strong WA in \iras, for which only a preliminary modeling was included in L15.

In the present work we focus our analysis in the WA in \iras. We report results on its physical properties, and characterize its variability between 2004 and 2014. 

\hspace{1cm}
\section{X-ray observations and data reduction}

\xmm\ observed \iras\ twice in 2004 for a total exposure time of 40~ks (August 30th, OBSID: 0206860101; and September 5th, OBSID: 0206860201) and two more times in 2014 for a total exposure time of 160~ks (January 23rd, OBSID: 0721220101; and January 25th, OBSID: 0721220301). There are no background flares during the observations. Standard data processing from the Reflection Grating Spectrometer \citep[RGS,][]{2001A&A...365L...7D} was carried out with the {\small{SAS v13.5.0}} tool {\it rgsproc}. 
In order to maximize the signal-to-noise,  RGS spectra of each epoch were combined in one single spectrum by using the {\small{SAS}} tool \textit{rgscombine}. This can be done because the flux variability within the two \xmm\ observations of each epoch is negligible, and there are no spectral changes between them ($\Delta\Gamma<5$\% and $\Delta$Flux=3\%, where $\Gamma$ is the continuum spectral slope). 

The software used to perform the spectral analysis is {\small{XSPEC}} v.12.10.0c \citep{1996ASPC..101...17A}. 

\hspace{1cm}
\section{Spectral analysis}

The spectral analysis was performed on the unbinned  RGS spectra, therefore the $C$-statistic \citep{1979ApJ...228..939C} was applied for the spectral fittings. Errors correspond to 1$\sigma$ level throughout this work. 
A $\Lambda$CDM cosmology is assumed, with H$_0=70$\,\kms\,Mpc$^{-1}$,
$\Omega_\Lambda= 0.73$, and $\Omega_M = 0.27$.

\hspace{1cm}
\subsection{2014 \xmm\ data} \label{sec:2014}

We start our analysis by considering the merged  RGS data, representative of the spectrum in 2014. 
Based on our thorough previous analysis of \iras\ (L15), we consider a complex model consisting of a power law \xr\ continuum modified by Galactic absorption \citep{2005A&A...440..775K} of column density $N_{\rm H}$=2$\times$10$^{20}$~cm$^{-2}$. The numerous narrow absorption lines imprinted by ionized gas to this continuum, were modeled by means of the self-consistent photoionization code {\small{PHASE}} \citep{2003ApJ...597..832K}. 
Each one of the {\small{PHASE}} components characterises one layer of ionized gas in terms of a set of parameters: 
(i) the ionization parameter, defined as the logarithm of $U = Q~(4 \pi n R^2 c)^{-1}$ \citep{1987MNRAS.225...55N, netzer2008}, being $Q$ the photon rate integrated over the entire Lyman continuum, $n$ the gas number density and $R$ the gas distance from the nuclear source of photons, 
(ii) the equivalent hydrogen column density, 
(iii) the outflow velocity, and 
(iv) the internal microturbulent velocity. 
The electron temperature in the {\small{PHASE}} model corresponds to the photoionization equilibrium
of the gas.
To compute the absorption spectrum of \iras , we used the same spectral energy distribution as in L15. 
Four {\small{PHASE}} components were required for modeling the WA, and five more for the  UFO (L15). 
The  UFO properties are exhaustively studied and characterised in the narrow band (18-23~\AA) in our previous work. 
In the following, the focus is set in modeling the four WAs in the entire 7-35~\AA~band of the  RGS spectrum. We initially kept all the five  UFO components of L15. However, only the three most significant ones ($\sigma > 3$ in Table 2 of our previous work) are recovered with equal weight in the entire 7-35 \AA\ band. 
This is due to the fact that various free parameters of the remaining two  UFO components found at lower significance in L15 are not sensitive enough to the broad-band data.
In addition, when considering the entire  RGS band, we note that one  UFO component of L15 (Comp.\,A) is better fitted by two different phases of gas outflowing at the same velocity (more detail in Section\,\ref{sec:consistency}).

The ionization, column density, and velocity of the four WAs and the four  UFOs are left free. The internal microturbulent velocity of the gas in the  UFO components is fixed to 50\,\kms , because this parameter is not sensitive to the data when the absorption lines are very narrow. 
The turbulent velocity of the WA is left free, together with the power law photon index and normalization. 

\begin{figure*}
\begin{center}
\mbox{\includegraphics[width=0.22\textwidth,angle=-90]{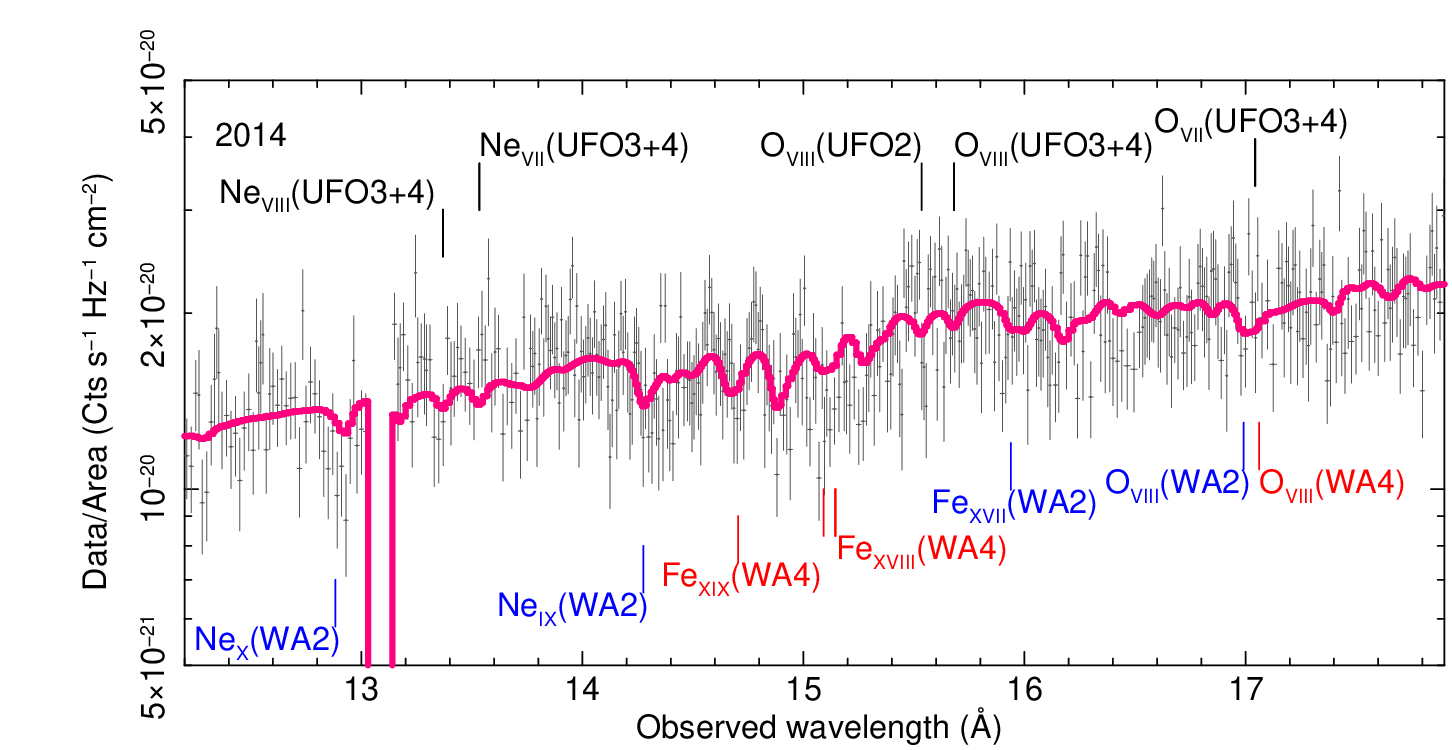}}
\mbox{\includegraphics[width=0.22\textwidth,angle=-90]{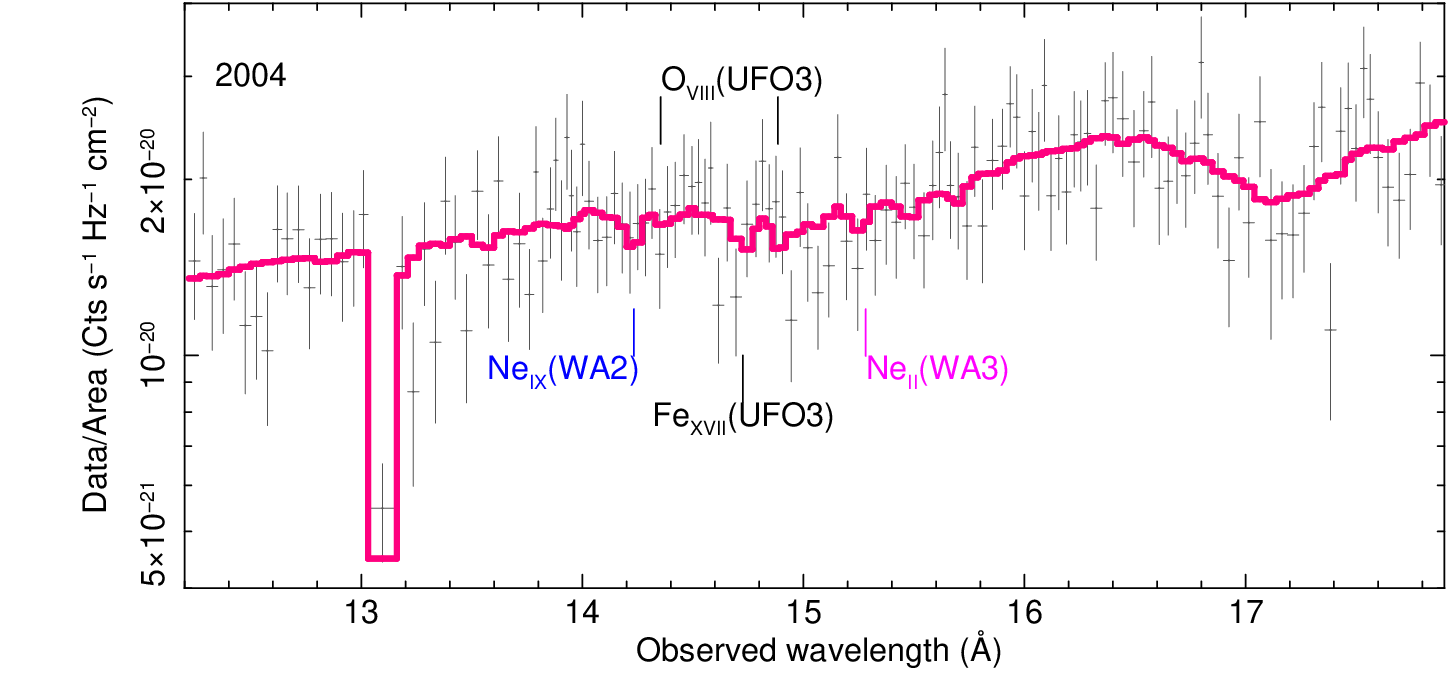}}
		
{\vspace{0.0cm}}

\mbox{\includegraphics[width=0.22\textwidth,angle=-90]{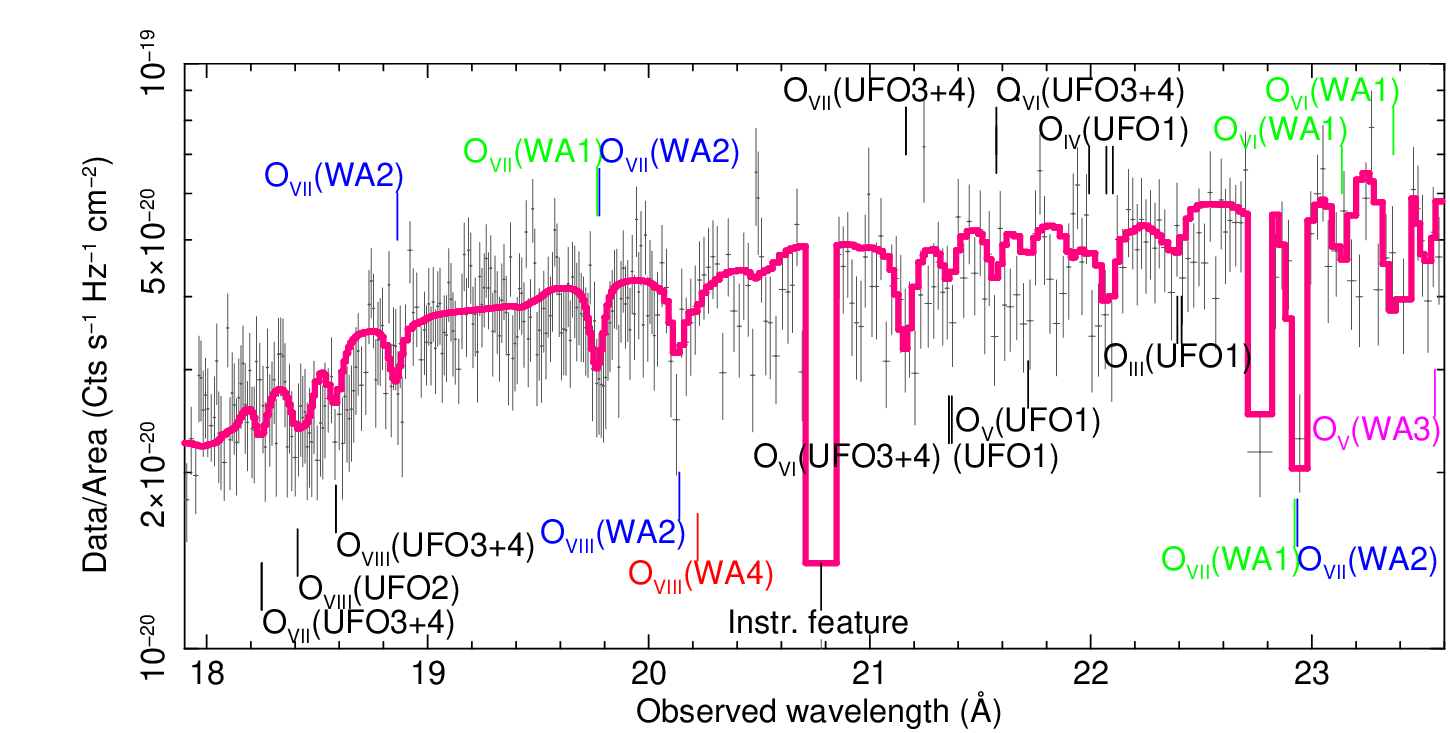}}
\mbox{\includegraphics[width=0.22\textwidth,angle=-90]{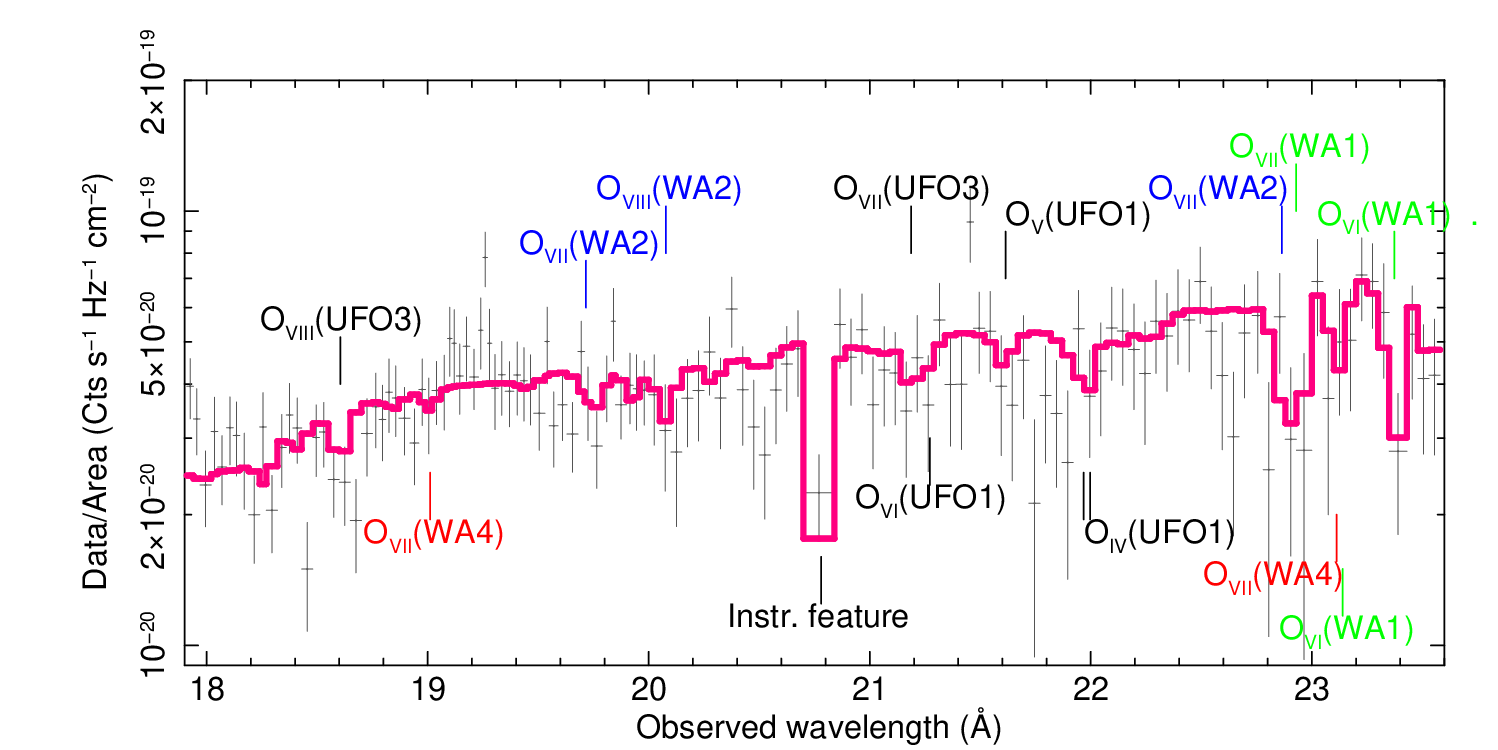}}

{\vspace{0.0cm}}

\mbox{\includegraphics[width=0.22\textwidth,angle=-90]{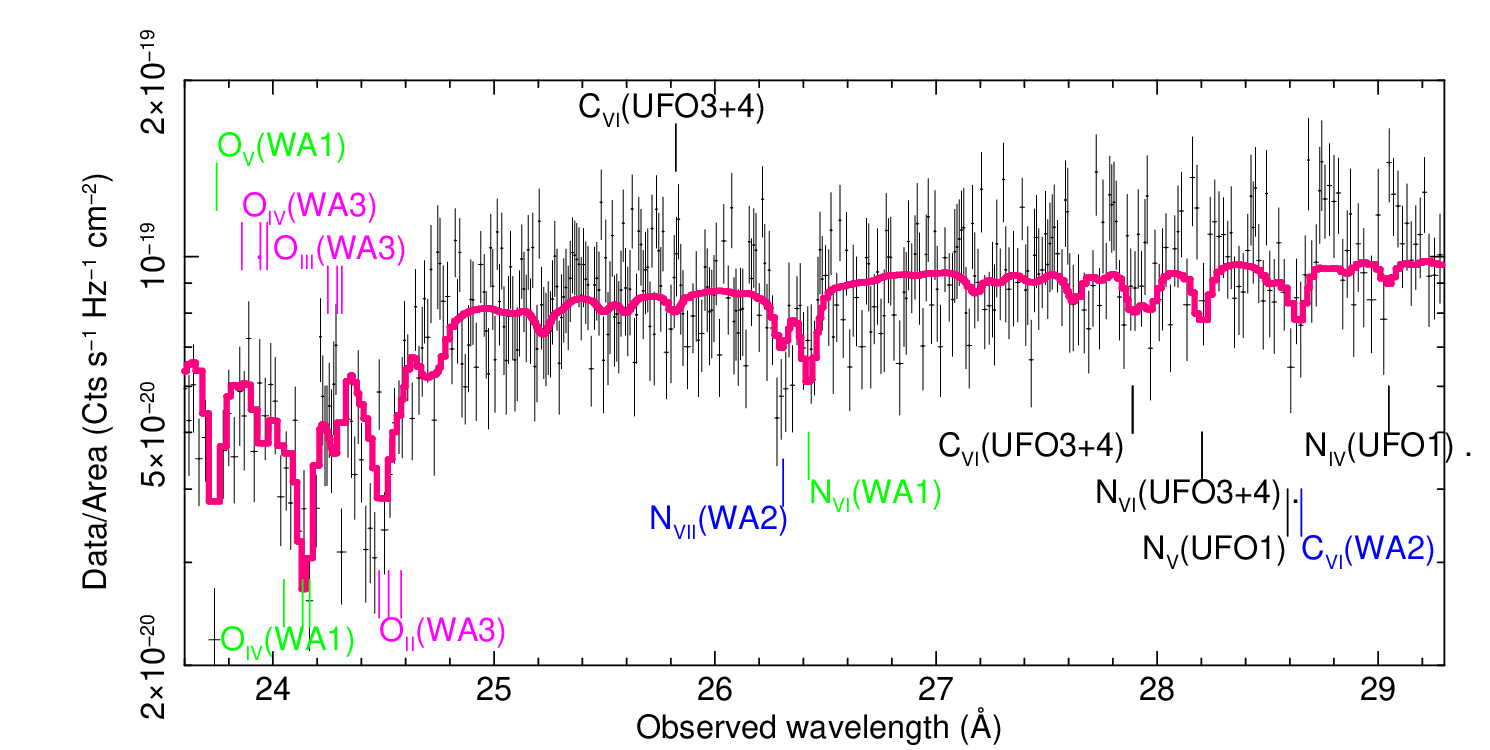}}
\mbox{\includegraphics[width=0.22\textwidth,angle=-90]{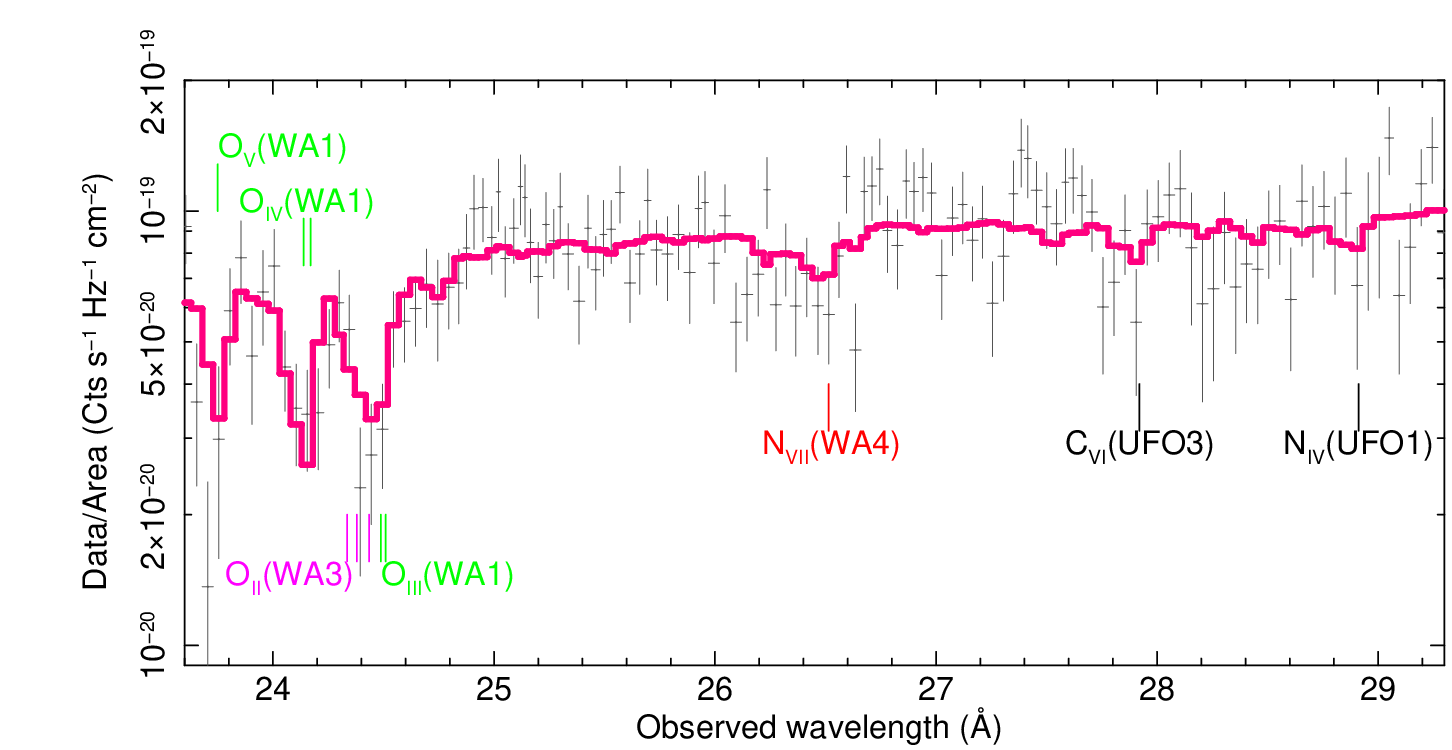}}

{\vspace{0.0cm}}

\mbox{\includegraphics[width=0.22\textwidth,angle=-90]{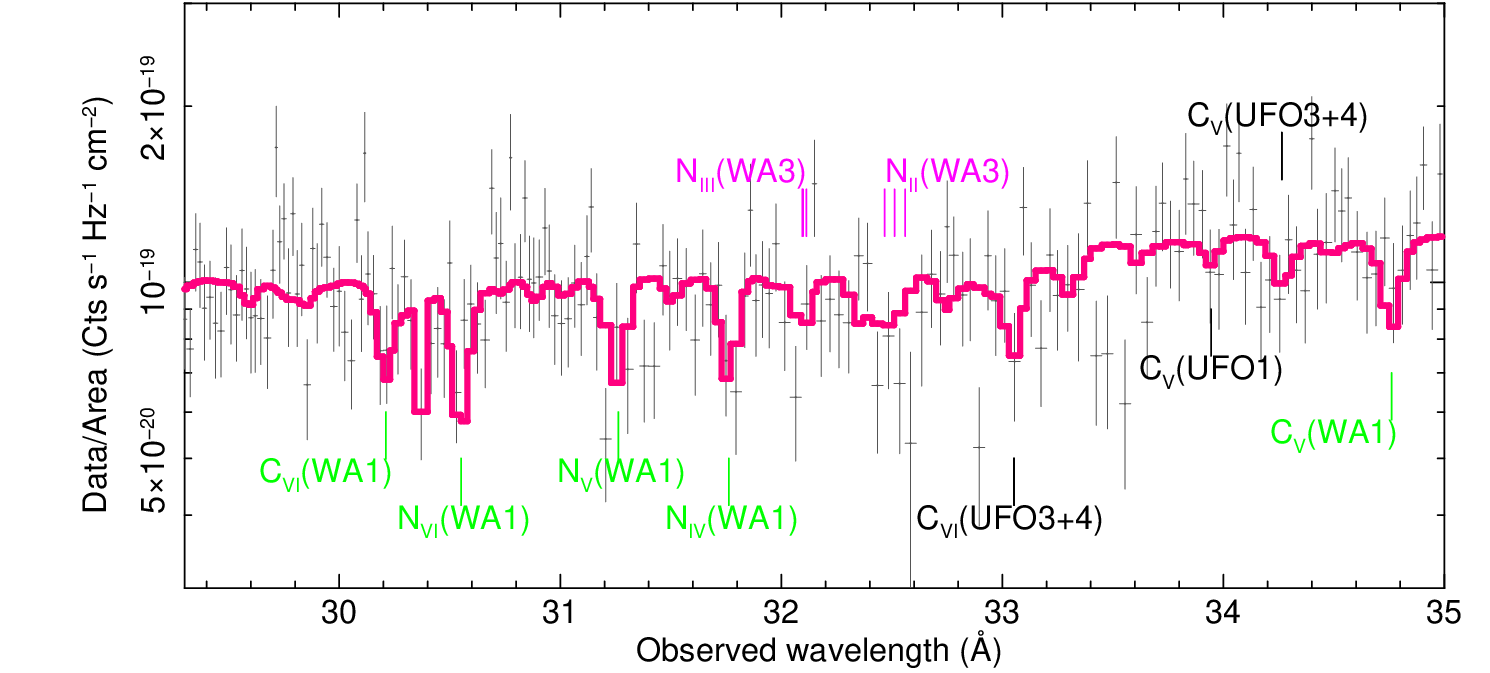}}
\mbox{\includegraphics[width=0.22\textwidth,angle=-90]{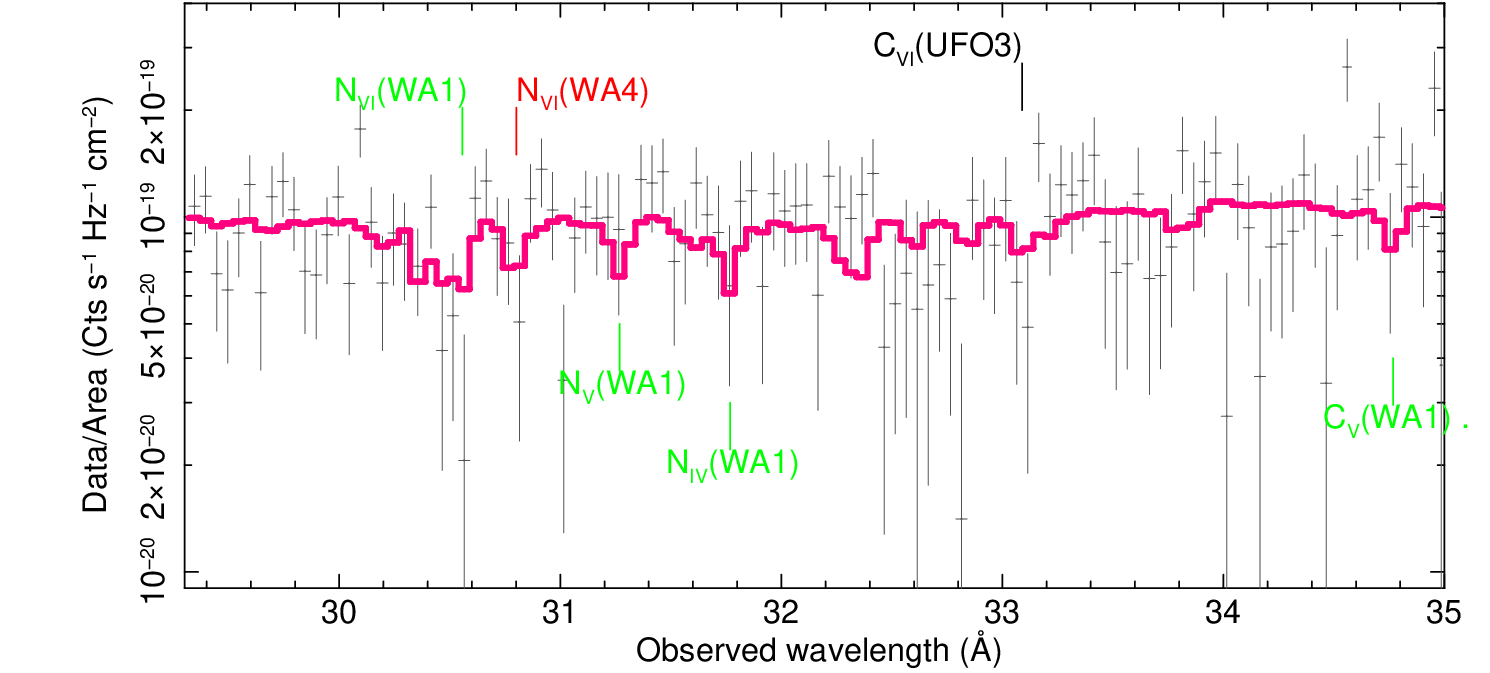}}

\caption{Detector area-corrected data and best-fitting model with a Galactic-absorbed continuum + 4 WAs + 4 (2)  UFOs applied to the 2014 (2004) spectrum on the left-hand (right-hand) panels. We show the entire broad band in four wavelength ranges, from top to bottom: 12.2-17.9~\AA , 17.9-23.6~\AA , 23.6-29.3~\AA , and 29.3-35~\AA . The main absorption lines produced by every component are plotted in different colors: WA~1 in green, WA~2 in blue, WA~3 in magenta, WA~4 in red, and the  UFOs in black. Every line marked in this figure is included in Table\,\ref{tab:lines2014}. 
\label{fig:model}}
\end{center}
\end{figure*}

The best-fitting {\small{PHASE}} parameters of the 4 WAs and 4  UFOs are shown in the left-hand side of Table~\ref{tab:wa2014}. 
We measure a soft flux level of 
$0.99^{+0.01}_{-0.02} \times 10^{-11}$\,erg\,cm$^{-2}$\,s$^{-1}$ 
fitted in the  RGS band 7-35~\AA , which is equivalent to the $\sim$\,0.3-2~keV range. The power law photon index is $\Gamma = 2.99^{+0.04}_{-0.01}$, typical of NLS1 galaxies in the soft \xr\ band.
We find one of the WAs to be inflowing with a velocity component along the line of sight of $1750 \pm 250$~\kms . 
The most significant line associated with this inflow is  O${\textsc{viii}}$ at 20.22~\AA , with an equivalent width (EW) of $\sim$~9~\AA\ (see component WA~4 in the left-hand panels of Fig.~\ref{fig:model} and on the left side of Table~\ref{tab:lines2014}). 
Two additional components (WA~1 and WA~2 in Table~\ref{tab:wa2014}) are inflowing too at few hundreds \kms , although their radial velocities are consistent with 0. That is so because the  RGS spectral resolution equals a few hundreds \kms\ at the wavelengths involved \citep{2001A&A...365L...7D}, 
and the accuracy of the energy scale equals 5~m\AA\ \citep{2015A&A...573A.128D}. 
The other WA, which is the least ionized, is instead outflowing at $2300 \pm 200$~\kms\ (WA\,3, see Table\,\ref{tab:wa2014}, left). The column densities are typical for WAs ($N_{\rm H}$\,$\sim$\,$10^{20}$\,-\,$10^{21}$~cm$^{-2}$), while the ionization parameters are in the low range ($\log{U}$\,$\sim$\,$-2.5$\,-\,$0.4$). 
The internal microturbulent velocities of the WAs are low, being the largest one $v_{\rm turb} = 160 \pm 30$\,\kms , which is reflected in narrow, unresolved absorption lines.

Our model provides a robust characterization of the WA and a good description of the data, and yields a statistics of $C$=3045 for 2738 degrees of freedom in the 7-35~\AA~band. The inclusion of an additional (fifth) absorber does not improve the fit in a statistical way. 

The statistical significance of each WA and  UFO component  is represented by the improvement of $\Delta C$ shown in the left side of Table~\ref{tab:wa2014}. 
We compute the statistical significance of a given absorption component by calculating the improvement in Cash statistics between the best-fit with the baseline model, and the best-fit obtained when that component is removed from the baseline model.

According to the fit parameters reported in the left side of Table~\ref{tab:wa2014}, the most significant components of the wind are WA\,1 ($\log U$\,$\sim$\,$-1.9$, $\log{N_{\rm H}}$\,$\sim$\,$21.1$), and WA\,2, with the same column density but higher ionization ($\log U$\,$\sim$\,$-0.6$), both of which are tracing gas consistent either with being at rest or inflowing at a few hundreds~\kms . 
WA\,3 is a cooler, fainter component ($\log U$\,$\sim$\,$-2.5$, $\log{N_{\rm H}}$\,$\sim$\,$20.9$), outflowing at $\sim$\,$2300$~\kms . 
The fourth and last of the WA components is the most ionized shell ($\log U$\,$\sim$\,$0.4$), and is inflowing at $\sim$\,$2 \times 10^3$~\kms . 

We now describe the model developed to fit the ultra fast wind in the present work. The  UFO here is still structured as a multi-phase, multi-component wind with two different velocities close to $\sim$\,$0.1c$. The first component is outflowing at $26900 \pm 200$\,\kms , and is formed by at least two different shells, named UFO\,1 and UFO\,2 in Table~\ref{tab:wa2014}. When we let the velocities of these two shells free to vary independently, we still get the same outflow velocity for both of them. 
UFO\,1 traces a cool and shallow component of the wind ($\log U = -2.47^{+0.15}_{-0.19}$, $\log{N_{\rm H}} = 20.10^{+0.06}_{-0.09}$). On the contrary, UFO\,2 is five orders of magnitude more ionized ($\log U = 2.63^{+0.04}_{-0.13}$) and three orders of magnitude denser ($\log{N_{\rm H}} = 23.70 \pm 0.15$).

The second  UFO component is formed by two shells as well, named UFO\,3 and UFO\,4 in Table~\ref{tab:wa2014}, which are outflowing at $24100 \pm 100$\,\kms . 
UFO\,3 is a cool, faint layer ($\log U = -0.35^{+0.12}_{-0.19}$, $\log{N_{\rm H}} = 20.4^{+0.2}_{-0.4}$), and UFO\,4 is cooler and slightly thicker ($\log U$\,$\sim$\,$-1.22^{+0.10}_{-0.05}$, $\log{N_{\rm H}}$\,$\sim$\,$20.85^{+0.03}_{-0.06}$).

As in the case of the UFO\,1\,\,+\,\,UFO\,2 component, when the velocities of UFO\,3 and UFO\,4 are allowed to vary independently, the same outflow velocity is recovered for the two shells. 
Each component of the flow is therefore identified by its outflow velocity. The need to fit the gas at the same velocity with 2 separate {\small{PHASE}} components is driven by the very wide range of ionization and column density spanned by the gas, that cannot be fitted by one single {\small{PHASE}} model. 

The features imprinted in the 2014 spectrum by the four WAs, as well as by the four  UFOs, can be seen in the left panels of Fig.~\ref{fig:model}. Their wavelengths and EWs are gathered on the left side of Table~\ref{tab:lines2014}.

%WIND COMPONENTS TABLE
\begin{deluxetable*}{l|ccccc|ccccc}
	\rotate
	\tablecaption{Parameters of the 4 WA components and 4  UFOs detected in the merged  RGS data representative of the 2014 (left) and 2004 (right) spectra. \label{tab:wa2014}}
	\tablecolumns{12}
	\tablewidth{0pt}
	\tablehead{
		\colhead{}    &  \multicolumn{4}{c}{2014} &   \multicolumn{4}{c}{2004} \\ 
		\colhead{Component} &
		\colhead{$\log{U}$} &
		\colhead{$\log{N_{\rm H}}$ \tablenotemark{(a)}} & 
		\colhead{$v_{\rm turb}$ \tablenotemark{(b)}} &  
		\colhead{$v$ \tablenotemark{(c)}} &  
		\colhead{$\Delta C$} &
		\colhead{$\log{U}$} &
		\colhead{$\log{N_{\rm H}}$ \tablenotemark{(a)}} & 
		\colhead{$v_{\rm turb}$ \tablenotemark{(b)}} &  
		\colhead{$v$ \tablenotemark{(c)}} &  
		\colhead{$\Delta C$}
	}
	\startdata
	WA\,1 (rest) & $-1.88 ^{+0.03}_{-0.02}$ & $21.09 ^{+0.01}_{-0.01}$ & $160 \pm 30$ & $-320 \pm 70$ & $530$ & $-2.10 ^{+0.07}_{-0.06}$ & $21.01 ^{+0.03}_{-0.11}$ & $170 \pm 60$ & $-380 \pm 160$ & $100$ \\
	WA\,2 (rest) & $-0.57 ^{+0.05}_{-0.04}$ & $21.12 ^{+0.03}_{-0.04}$ & $<40$ & $-430 \pm 90$ & $120$ & $-0.57 ^{+0.08}_{-0.11}$ & $21.25 ^{+0.11}_{-0.20}$ & $<50$ & $-490 \pm 110$ & $61$ \\
	WA\,3 (outflow) & 	$-2.47 \pm 0.02$ & $20.93 \pm 0.01$ & $<40$ & $2300 \pm 200$ & $90$ & $-2.81 ^{+0.07}_{-0.19}$ & $20.88 ^{+0.03}_{-0.03}$ & $<60$ & $4000 \pm 200$ & $34$ \\ 
	WA\,4 (inflow) & $0.35 ^{+0.11}_{-0.16}$ & $20.84 ^{+0.16}_{-0.14}$ & $100 \pm 60$ & $-1750 \pm 250$ &  $24$ & $-1.3 ^{+0.3}_{-0.4}$ & $20.6 ^{+0.2}_{-0.4}$ & $110 \pm 70$ &$-2900 \pm 200$ & $14$ \\ 
	UFO\,1 & $-2.47 ^{+0.15}_{-0.19}$ & $20.10 ^{+0.06}_{-0.09}$ & $50^{\rm f.}$ & $26900 \pm 200$ & $27$ & $-2.2 ^{+0.3}_{-0.2}$ & $20.4 \pm 0.2$ & $50^{\rm f.}$ & $28500 \pm 1000$ & $9$ \\
	UFO\,2 & $2.63 ^{+0.04}_{-0.13}$ & $23.70 \pm 0.15$ & $50^{\rm f.}$ & '' & $7$ & $-$ & $-$ & $-$ & $-$ & $-$ \\
	UFO\,3 & $-0.35 ^{+0.12}_{-0.19}$ & $20.4 ^{+0.2}_{-0.4}$ & $50^{\rm f.}$ & $24100 \pm 100$ & $5$ & $-0.30 ^{+0.05}_{-0.07}$ & $21.27 ^{+0.15}_{-0.19}$ & $50^{\rm f.}$ & $23900 \pm 100$ & $36$ \\
	UFO\,4 & $-1.22 ^{+0.10}_{-0.05}$ & $20.85 ^{+0.03}_{-0.06}$ & $50^{\rm f.}$ & '' & $4$ & $-$ & $-$ & $-$ & $-$ & $-$ \\
	\enddata
	\tablenotetext{a}{$N_{\rm H}$ in cm$^{-2}$.}
	\tablenotetext{b}{Microturbulent velocity of the gas in \kms . }
	\tablenotetext{c}{Velocity component along the line of sight in \kms . Negative / positive values respectively refer to inflowing / outflowing material. }
	\tablecomments{
		Fit to 2014 data (left): $C/dof = 3045 / 2738 $, 
		$\Gamma = 2.99 ^{+0.04}_{-0.01}  $; 
		Fit to 2004 data (right): $C/dof = 2848 / 2748 $, 
		$\Gamma = 2.99 ^{+0.17}_{-0.07} $. 
	}
\end{deluxetable*}

\subsection{2004 \xmm\ data} \label{sec:variability}

We consider now the merged  RGS spectrum of the two data sets obtained  ten years earlier, in 2004. In our previous work, we inspected the possible presence of a fast wind by applying the best fit model of 2014 comprising 5~ UFOs. One of these  UFO components is confirmed in the 2004 spectrum, and the others cannot be statistically ruled out (see L15 for more detail). 
Considering the larger spectral band, we recover two of those  UFO components, outflowing at $28500 \pm 1000$\,\kms\ and at $23900 \pm 100$\,\kms , respectively (see right side of Table~\ref{tab:wa2014}). Hence, in order to check for variability at long timescales, we apply a model consisting of a locally absorbed power law plus four WAs plus two  UFOs to the  RGS data from 2004.
We compute a flux level of $1.03^{+0.01}_{-0.08}\times 10^{-11}$\,erg\,cm$^{-2}$\,s$^{-1}$ in the 0.3-2\,keV range, therefore flux variability of the source in a 10-y lapse is minor. The results of this fit are shown in the right side of Table~\ref{tab:wa2014}. 

WA components no.~1 and 2 are consistent in 2004 and 2014, so that it is conceivable that both of them remain at rest (or keep inflowing at few hundreds \kms ) after a decade. 
Wind parameters of WA\,3 are not consistent in 2004 and 2014; specifically, the large difference in the velocity measured in the two epochs suggests that this WA in 2004 and 2014 is most likely originated by two different blobs of gas, both of which are escaping from the system at $4000 \pm 200$ and $2300 \pm 200$~\kms , respectively. 
As for WA\,4, we observe marginal consistency between its column densities in 2004 and 2014, though its ionization in 2014 is between one and two orders of magnitude larger than in 2004; also, this component is still inflowing, but at a lower velocity in 2014 ($-1750 \pm 250$~\kms , to be compared with $-2900 \pm 200$~\kms\ ten years earlier). 

As for the  UFOs, we recover the two components at different velocities in the 2004 data, but their multi-phase nature can't be detected (nor can be statistically ruled out). Both UFO\,1 and UFO\,3 remain remarkably unvaried within a decade, perhaps indicating the presence of two persistent  UFO components (see Table\,\ref{tab:wa2014}). 

We interpret the behavior of WA components no.~3 and 4 in the two observations as the result of the appearance and disappearance of different wind's sections, while WA components no.~1 and 2 and  UFO components 1 and 3 are persistent over a decade (see Section~\ref{sec:discussion}). 
The features imprinted in the 2004 spectrum by the two  UFO and the four WA components can be seen in the righ-hand panels of Fig.~\ref{fig:model}, and are gathered in the right-hand side of Table~\ref{tab:lines2014}. 

We offer a detailed interpretation of the results from the spectral analysis in Section~\ref{sec:discussion}.

\subsection{Consistency with previous results} \label{sec:consistency}
\hspace{1cm}

In our previous work on the winds system in \iras , five  UFOs were detected and characterized from 2014 \xmm\ data in the 18-23\,\AA\ narrow band (see Table\,2 in L15). In addition, one of those  UFOs was found to be persistent on a 10-yr scale. 
In this paper we perform a thorough analysis over the 7-35\,\AA\ broad band, recovering three out of the five outflows previously reported. The reason why the two less significant components cannot be constrained in the broad band is because the most prominent spectral features associated with such components are three absorption lines, as reported in Table\,1 in L15: 
O${\textsc{viii}}$ at $18.60 \pm 0.03$, 
O${\textsc{iii}}$ at $21.83^{+0.03}_{-0.01}$, and 
O${\textsc{iv}}$ at $21.55^{+0.03}_{-0.02}$. 
These lines are significant enough in the 18-23\,\AA\ narrow band so that the model parameters of the components producing them in L15 are sensitive to the data in this band. However, in the 7-35\,\AA\ broad band their effect gets overwhelmed by the other absorption features. This way, we can neither confirm the presence of the two less significant  UFO components, nor rule-out their existence. 

The three  UFO components that are recovered in this paper are those called ``A'', ``B'', and ``C'' in L15. Component ``A'' is split here in two phases, namely UFO\,3 and UFO\,4. Components ``B'' and ``C'' are depicted in this work, as in L15, as a single outflow with two phases, UFO\,1 and UFO\,2, as described in Section\,\ref{sec:2014}.

As of the 2004 data, two  UFO components are recovered. UFO\,3, is detected with a significance level of $\Delta C = 36$. This component was already detected in our previous work, where the rest of the  UFOs couldn't be statistically ruled out. In this paper, we find one of those extra components: UFO\,1, with a low significance level of $\Delta C = 9$. 

\hspace{1cm}
\section{Discussion} \label{sec:discussion}
\hspace{1cm}

The \xmm\ high resolution spectra of \iras\ have revealed an outflow characterized by very complex behavior. The source presents the simultaneous presence of a stratified  UFO and a multi-layered slower absorber whose components are flowing in and outward. The availability of two spectral epochs allows us to track the evolution of the slower wind along the 10 years elapsed in between the 2 \xmm\ observations. Notably, the  UFO does not present any hint of evolution between the two epochs, although this can be said only with respect to the two persistent components that are seen in 2004 and 2014. The following discussion is based on both results from our 2015 paper and from the present work. 

\hspace{1cm}
\subsection{The source: properties of \iras\ } \label{sec:source}
\hspace{1cm}

We firstly attempt to set some spatial scales and distances for the wind. 
The launch radii of the outflow component in 2004 and 2014 were estimated by assuming that the outflow velocity must be larger than (or equal to) the escape velocity at the launch radius, which is computed as $R=\frac{2GM_{\rm BH}}{v^2_{\rm esc}}$ by definition. Considering the outflow velocities measured for WA\,3, which is the only WA component outflowing in both  epochs, the launch radius must be greater than $8.4 \times$10$^{15}$~cm in 2004 and than $2.5 \times$10$^{16}$~cm ten years later, in 2014. 

Following \cite{2006ApJ...648L.101E}, we compute the dust sublimation radius in this object as $R_{\rm dust} = 0.4 L^{0.5}_{45}$\,pc, where $L_{45}$ is the bolometric luminosity of the source in units of $10^{45}$\,erg\,s$^{-1}$. Hence, $R_{\rm dust} = 0.3$\,pc = $9\times 10^{17}$\,cm, \ie\ the dust sublimation radius is roughly two orders of magnitude larger than the launch radius of the outflow component, therefore it is likely for the outflow material to be dust-free. With this in mind, together with the results exposed in Sections~\ref{sec:2014} and \ref{sec:variability}, in the following we investigate three possible interpretations of the nature of the absorbing material detected in \iras , as well as to its dynamics.

\hspace{1cm}
\subsection{Shocked outflow interpretation} \label{sec:shock}
The detection of several components of gas flowing inward and outward at different velocities with the variability pattern observed during the two \xmm\ visits, suggests a different nature of this wind compared to ``standard'' WAs. Indeed, it is difficult to imagine that such a composite flow of ionized gas resides in external regions like the NLR of the AGN. 

The coincidence of the  UFO with slower winds moving in opposite directions may be explained in terms of a ``shocked outflow.''
This model predicts that an initial fast outflow  radiatively launched at accretion disk scale with outflow velocity  $v_{\rm out}$~$\ge$ 10$^4$~\kms\ shocks with the ambient medium \citep{2010MNRAS.402.1516K, 2012MNRAS.425..605F, 2012ApJ...745L..34Z, 2015ARA&A..53..115K}.  This impact produces first a reverse shock of the wind with the gas located at a radius where the escape velocity is lower than the outflow velocity, therefore the wind keeps sweeping up the surrounding material and develops a second forward shock. The two shock fronts are separated by a contact discontinuity and whereas the shocked ambient gas could decelerate to velocity of the order of 100~\kms, the wind shock (reverse) maintains its high velocity while entraining the ambient gas and pushing it further out \citep{2012MNRAS.425..605F}. 
In the case of the Narrow Line Seyfert Galaxy NGC~4051, \cite{2011MNRAS.413.1251P}  also interpreted the signatures of the in-outflowing absorber in that system as the result of a shocked outflow, despite the non-detection of a nuclear ultra-fast wind.

\subsection{Instabilities in a shocked outflow?} \label{sec:instabilities}

As shown with much higher detail for the case of Supernova Remnants \citep{1998A&A...334.1060V}, at the discontinuity between the two shock fronts it is very likely that fluid instabilities (\eg\ Rayleigh-Taylor) start developing, since the densities of the impacting wind and of the impacted medium are considerably different. A condition for the Rayleigh-Taylor instability to grow is that the mass of the Interstellar Medium that is pushed by the discontinuity is higher than the mass of the ejecta \citep{1998A&A...334.1060V}, which undergoes a deceleration process that is able to trigger instabilities in the fluid. Such instabilities would easily alter the dynamics of the shocked outflow at the scale of the reverse shock and of the discontinuity, and they may give rise to slower components of the wind. It is conceivable that these blobs of gas at different conditions (velocity, ionization, density, temperature) would be continuously replenished by the effect of the instability and turbulence, and that they may also fall backwards instead of following the bulk of the outflow.
This behaviour is reproduced in simulations of a shocked outflow, where instabilities form turbulent plumes in the front side of the shocked material that reach velocities consistent with those measured in the  UFO multi-components \citep[][and Longinotti, Velazquez et al. in prep.]{2018arXiv180801043L}.
Depending on which plumes cross our line of sight, absorption lines corresponding to a wide range of velocity components will imprint an emerging absorption spectrum consisting of the combination of several gas components in outflow\,/\,inflow plus a stationary wind. 
Thus, the scenario that we are witnessing in the \xr\ spectra of \iras\ seems very consistent with the hypothesis of the shocked outflow described above.
In fact, we note that given the spectral resolution of the  RGS detectors, the slow inflow velocity measured for WA components no.~1 and 2 may actually indicate that these shells are stationary. 
The ionization state and velocity of these gas layers have not changed significantly along the 10-yr time scale explored by \xmm .   
With all probability this components represent the shock itself, whose velocity and likely position neither have changed significantly in the 10 years elapsed between the observations. 

In this case, the shock cannot overcome gravity and the outflow is in its momentum-conserving phase \citep[see \eg][]{2015ARA&A..53..115K}. 
On top of these stationary WAs, we observed significant changes in the velocity of the other 2 WA components: the apparent deceleration of $\sim$1000~\kms\ measured in components no.~3 and 4 between 2004 and 2014. 
It is likely that we are seeing different sections of gas rather than the same component changing its velocity. Remarkably,  these two components did change their ionization state from 2004 to 2014 (especially WA component no. 4), and since no change of the ionizing continuum is observed 
(the central source seems completely persistent within the two epochs), 
we need to invoke other mechanisms to explain the variety of ionization states present in the flowing gas.
A viable explanation for this behaviour is to postulate that these short-lived blobs\,/\,sections of slow gas are produced in a turbulent regime upon a shocked outflow. 
The effect of the Rayleigh-Taylor instability due to the different densities of the pre- and post-shock gas is responsible for driving them in and out of our line-of-sight. 

%It is worth of mention to note that more recent deep observations performed by \textit{Chandra} LETG in late 2016-early 2017 corroborate the proposed scenario since they confirmed  both the persistency of the fast wind seen by \xmm\ in \iras\ and the presence of a variable multi-component WA fully consistent with the observational properties above described  (Longinotti et al. in prep.). 

In addition, we also note that shocked outflow models predict that synchrotron radiation is emitted in the core of radio-quiet AGN due to shock processes \citep{2015MNRAS.447.3612N, 2014MNRAS.442..784Z}. Recent results from very long baseline interferometry (VLBI) observations of \iras\ reported by \cite{2017A&A...600A..87G}, although not conclusively, show that  the compact synchrotron  emission detected in this source may indicate an origin in a shocked outflow. New deep VLBI observations recently obtained will allow us to gain insights on the association of the radio emission to a shocked outflow (Giroletti et al. in prep).

\hspace{1cm}
\subsection{Alternative interpretations}

Other interpretations of the observed outflow pattern are possible and we explore them in this section.  
The main observational result that shall be addressed is a valid mechanism of producing different sections\,/\,blobs of gas at different ionization states without variations in their ionization being produced by the steady central photoionization source. 
We are therefore inclined to consider scenarios where clouds or clumps of gas are involved, as it would be difficult to explain the change in ionization and velocity observed over the 10~yr lapse for a continuous distribution of gas.

\cite{2017ApJ...847...56E} proposes that the warm, radiation pressure driven wind from the accretion disk in AGN and quasars is the material from which the dense, cool clouds in the BLR are formed, by means of condensation before the WA outflows can reach the escape velocity of the system. Those condensed clouds, unlike the WA, cannot gain acceleration enough to reach the escape velocity. Therefore, they ``rain'' toward the SMBH as an inflow of short-lived clouds. 
With respect to our results, these raining clouds could be two different stages of what we called WA components no.~3 and 4, while no.~1 and 2 would be the stable, extended and less dense components from which the others are condensed. 

Another possible interpretation is provided by the condensed clouds scenario, which involves no shock. Instead, the inflow occurs via chaotic cold accretion: cold clouds ``rain'' from the material that cools as it flows away from the launching region \citep[see e.g.][]{2017ApJ...837..149G}. Components that do not reach the escape velocity for this system simply fall down again. 

However, the condensed clouds scenario cannot explain the ionization of some of the components in this system. In fact, we do not detect cold gas at all, but warm instead ($T > 10^4$~K). Finally, considering that free-fall velocities computed for this system down to the distances where the WAs are located are roughly $\sim$\,$400$\,\kms, we conclude that neither of the two alternative scenarios can explain inflow velocities of the order of $2-3\times 10^3$ \kms\ like the ones  detected in the \xmm\ spectra. 

Therefore, both alternative interpretations fail to provide a consistent explanation of the whole ensemble of observational results derived from \iras\ spectra, leaving the shocked outflow model framework as the most likely scenario for this peculiar wind system.

\acknowledgements 
This paper is based on observations obtained with \xmm , an ESA science mission with instruments and contributions directly funded by ESA Member States and NASA. 
The authors wish to thank the anonymous referee for a exhaustive review and useful comments that helped to improve the text. 
We acknowledge Dr. Pablo Velázquez for his valuable contribution in Section\,\ref{sec:instabilities}. 
MS thanks UNAM for support through the DGAPA Program for Post-Doctoral Scholarships. 
YK acknowledges support from the grant PAIIPIT IN106518. 
ALL and YK acknowledge financial support from the ESAC Faculty.

%%%%%%%%%%%%%%%%%%%%%%%%%%%%%%%%%%%%%%%%%%%%%%%%%%%%%%%%%%%%%%%%%%%%%%%%
%% BIBLIOGRAPHY

%%%%%%%%%%%%%%%%%%%%%%%%%%%%%%%%%%%%%%%%%%%%%%%%%%%%
%%APÉNDICE

\startlongtable
\begin{deluxetable}{lccr|cr|cr}
	\tablecaption{Absorption lines produced by the absorbing components as identified with the {\small{PHASE}} model in \\the 2014 and 2004 \xmm\  RGS spectra. \label{tab:lines2014}}
	\tablecolumns{7}
	\tablewidth{0pt}
	\tablehead{
		\multicolumn{4}{c}{ } & \multicolumn{2}{c}{2014} & \multicolumn{2}{c}{2004} \\ 
		\colhead{Ion} &
		\colhead{Transition} &
		\colhead{$\lambda$ (\AA) \tablenotemark{(a)}} & 
		\colhead{Abs.} &
		\colhead{Obs. $\lambda$ (\AA) \tablenotemark{(b)}} & 
		\colhead{EW} (\AA) &
		\colhead{Obs. $\lambda$ (\AA) \tablenotemark{(b)}} & 
		\colhead{EW} (\AA) 
	}
	\startdata
	Ne${\textsc{x}}$       & $1s^1 \rightarrow 2p^1$ (K$\alpha$)                &  $12.134$  & WA\,2   & $12.882 \pm 0.001$          &  $4 \pm 1$      &  $ $                         &  $ $  \\
	Ne${\textsc{ix}}$      & $1s^2 \rightarrow 1s^1 2p^1$ (K$\alpha$)           &  $13.447$  & WA\,2   & $14.276 \pm 0.001$          &  $15 \pm 1$     &  $14.233 \pm 0.001$          &  $17^{+1}_{-2}$ \\
	Ne${\textsc{viii}}$    & $1s^2 2s^1 \rightarrow 1s^1 2s^1 2p^1$ (K$\alpha$) &  $13.646$  & UFO\,3  & $13.369 \pm 0.001$          & $8 \pm 1$       &  $ $                         &  $ $  \\
	Ne${\textsc{viii}}$    & '' &  $13.646$  & UFO\,4  & $13.370 \pm 0.001$          & $7 \pm 1$       &  $ $                         &  $ $  \\
	Fe${\textsc{xix}}$     & $2p^4 \rightarrow 2p^3 3d^1$ (L$\alpha$)           &  $13.795$  & WA\,4   & $14.704^{+0.001}_{-0.008}$          &  $2 \pm 1$  &  $ $                         &  $ $  \\
	Ne${\textsc{vii}}$     & $2s^2 \rightarrow 1s^1 2s^2 2p^1$      (K$\alpha$) &  $13.814$  & UFO\,3  & $13.534 \pm 0.001$          & $13 \pm 3$      &  $ $                         &  $ $  \\
	Ne${\textsc{vii}}$     & '' &  $13.814$  & UFO\,4  & $13.534 \pm 0.001$          & $12 \pm 3$      &  $ $                         &  $ $  \\
	Fe${\textsc{xviii}}$   & $2p^5 \rightarrow 2p^4 3d^1$ (L$\alpha$) &  $14.158$  & WA\,4   & $15.092 \pm 0.001$          &  $2 \pm 1$  &  $ $                         &  $ $  \\
	Fe${\textsc{xviii}}$   & '' &  $14.208$  & WA\,4   & $15.144^{+0.001}_{-0.008}$  &  $4 \pm 1$      &  $ $                         &  $ $  \\
	Fe${\textsc{xviii}}$   & '' &  $14.208$  & WA\,4   & $15.144^{+0.001}_{-0.008}$  &  $ 4 \pm 1$  &  $ $                         &  $ $  \\
	Fe${\textsc{xviii}}$   & '' &  $14.256$  & WA\,4   & $15.196^{+0.001}_{-0.009}$          &  $2 \pm 1$      &  $ $                         &  $ $  \\
	Ne${\textsc{ii}}$      & $2p^5     \rightarrow 1s^1 2s^2 2p^6$  (K$\alpha$)  &  $14.600$  & WA\,3   & $ $                         &  $ $            &  $15.282 \pm 0.004$          &  $7^{+1}_{-2}$   \\
	O${\textsc{viii}}$     & $1s^1 \rightarrow 6p^1$ &  $14.634$  & UFO\,3  & $  $                        &  $  $           &  $14.35 \pm 0.01$          &  $5 \pm 3$ \\
	Fe${\textsc{xvii}}$    & $2p^6 \rightarrow 2p^5 3d^1$ (L$\alpha$) &  $15.014$  & UFO\,3  & $ $                         &  $ $            &  $14.73 \pm 0.01 $         &  $15 \pm 10$  \\
	Fe${\textsc{xvii}}$    & '' &  $15.014$  & WA\,2   & $15.939 \pm 0.001$          &  $6^{+3}_{-2}$  &  $ $                         &  $ $  \\
	O${\textsc{viii}}$     & $1s^1 \rightarrow 4p^1$ &  $15.176$  & UFO\,3  & $ $                         &  $ $            &  $14.89 \pm 0.01 $         &  $8 \pm 5$   \\
	O${\textsc{viii}}$     & $1s^1 \rightarrow 3p^1$ (K$\beta$) &  $16.006$  & UFO\,2  & $15.535 \pm 0.002$          &  $3^{+6}_{-2}$  &  $ $          &  $ $  \\
	O${\textsc{viii}}$     & ''  &  $16.006$  & UFO\,3  & $15.681 \pm 0.001 $         &  $7^{+1}_{-3}$      &  $ $                         &  $ $        \\
	O${\textsc{viii}}$     & ''  &  $16.006$  & UFO\,4  & $15.682 \pm 0.001 $         &  $7 \pm 1$      &  $ $                         &  $ $  \\
	O${\textsc{viii}}$     & ''  &  $16.006$  & WA\,2   & $16.992 \pm 0.001$          &  $5 \pm 1$      &  $ $                         &  $ $ \\  
	O${\textsc{viii}}$     & ''  &  $16.006$  & WA\,4   & $17.061^{+0.001}_{-0.010}$          &  $3 \pm 1$      &  $ $                         &  $ $  \\
	O${\textsc{vii}}$      & $1s^2 \rightarrow 1s^1 5p^1$ &  $17.396$  & UFO\,3  & $17.043 \pm 0.001$          &  $8 \pm 5$      &  $ $                         &  $ $  \\
	O${\textsc{vii}}$      & '' &  $17.396$  & UFO\,4  & $17.044 \pm 0.001$          &  $8 \pm 1$      &  $ $                         &  $ $  \\
	O${\textsc{vii}}$      & $1s^2 \rightarrow 1s^1 4p^1$ &  $17.768$  & WA\,2   & $18.863 \pm 0.001$          &  $3 \pm 1$  &  $ $                         &  $ $  \\
	O${\textsc{vii}}$      & '' &  $17.768$  & WA\,4   & $ $                         &  $ $            &  $19.01 \pm 0.02$          &  $11 \pm 5$  \\
	O${\textsc{vii}}$      & $1s^2 \rightarrow 1s^1 3p^1$ (K$\beta$) &  $18.627$  & UFO\,3  & $18.249 \pm 0.001$          &  $9^{+6}_{-7}$     &  $ $                         &  $ $  \\
	O${\textsc{vii}}$      & '' &  $18.627$  & UFO\,4  & $18.250 \pm 0.001$          &  $14 \pm 1$     &  $ $                         &  $ $  \\
	O${\textsc{vii}}$      & '' &  $18.627$  & WA\,1   & $19.769 \pm 0.001$          &  $32 \pm 1$     &  $ $                         &  $ $ \\
	O${\textsc{vii}}$      & '' &  $18.627$  & WA\,2   & $19.775 \pm 0.001$          &  $7 \pm 1$      &  $19.716 \pm 0.001$  &  $9 \pm 1$ \\
	O${\textsc{viii}}$     & $1s^1 \rightarrow 2p^1$ (K$\alpha$) &  $18.969$  & UFO\,2  & $18.411 \pm 0.002$          &  $17 \pm 5$     &  $ $                         &  $ $  \\
	O${\textsc{viii}}$     & '' &  $18.969$  & UFO\,3  & $18.584 \pm 0.001$          &  $17^{+2}_{-5}$     &  $18.61 \pm 0.02$          &  $20 \pm 13$  \\
	O${\textsc{viii}}$     & '' &  $18.969$  & UFO\,4  & $18.585 \pm 0.001$          &  $17 \pm 1$     &  $ $                         &  $ $  \\
	O${\textsc{viii}}$     & '' &  $18.969$  & WA\,2   & $20.138 \pm 0.001$          &  $26 \pm 1$     &  $20.078 \pm 0.001$  &  $28 \pm 2$ \\  
	O${\textsc{viii}}$     & '' &  $18.969$  & WA\,4   & $20.220^{+0.001}_{-0.012}$          & $9 \pm 1$  &  $ $                         &  $ $ \\
	O${\textsc{vii}}$      & $1s^2 \rightarrow 1s^1 2p^1$ (K$\alpha$) &  $21.601$  & UFO\,3  & $21.163 \pm 0.001$          &  $22^{+24}_{-14}$     &  $21.19 \pm 0.02$          &  $18 \pm 12$ \\
	O${\textsc{vii}}$      & '' &  $21.601$  & UFO\,4  & $21.163 \pm 0.001$          &  $46^{+2}_{-3}$     &  $ $                         &  $ $ \\
	O${\textsc{vii}}$      & '' &  $21.601$  & WA\,1   & $22.925 \pm 0.001$          &  $60 \pm 2$     &  $22.928^{+0.001}_{-0.007}$          &  $60^{+8}_{-16}$ \\
	O${\textsc{vii}}$      & '' &  $21.601$  & WA\,2   & $22.932 \pm 0.001$          &  $32 \pm 3$     &  $22.864 \pm 0.001$          &  $37^{+5}_{-7}$ \\
	O${\textsc{vii}}$      & '' &  $21.601$  & WA\,4   &  $ $                        &  $ $            & $23.112 \pm 0.001$           &  $22 \pm 15$ \\ 
	O${\textsc{vi}}$       & '' &  $21.800$  & UFO\,3  & $21.358 \pm 0.001$          &  $7 \pm 2$      &  $ $                         &  $ $  \\
	O${\textsc{vi}}$       & '' &  $21.800$  & UFO\,4  & $21.358 \pm 0.001$          &  $7^{+2}_{-1}$      &  $ $                         &  $ $  \\
	O${\textsc{vi}}$       & '' &  $21.800$  & WA\,1   &  $23.136 \pm 0.001$         &  $31 \pm 1$     & $23.140^{+0.001}_{-0.008}$           &  $32^{+2}_{-9}$ \\
	O${\textsc{vi}}$       & $1s^2 2s^1 \rightarrow 1s^1 2s^1 2p^1$ (K$\alpha$) &  $22.019$  & UFO\,1  & $21.371 \pm 0.001$          &  $9^{+3}_{-5}$      &  $21.27 \pm 0.01$          &  $10 \pm 8$  \\ %KELLY DB - refs 795 and 812
	O${\textsc{vi}}$       & '' &  $22.019$  & UFO\,3  & $21.573 \pm 0.004$          &  $20 \pm 3$     &  $ $                         &  $ $ \\
	O${\textsc{vi}}$       & '' &  $22.019$  & UFO\,4  & $21.573 \pm 0.001$          &  $19^{+3}_{-2}$     &  $ $                         &  $ $ \\
	O${\textsc{vi}}$       & '' &  $22.019$  & WA\,1   & $23.369 \pm 0.001$          &  $59 \pm 1$     &  $23.372^{+0.001}_{-0.007}$          &  $60^{+6}_{-8}$ \\
	O${\textsc{v}}$        & $2s^2     \rightarrow 1s^1 2s^2 2p^1$  (K$\alpha$)  &  $22.374$  & UFO\,1  & $21.716 \pm 0.001$          &  $16^{+1}_{-3}$     &  $21.61 \pm 0.01$          &  $13 \pm 9$  \\
	O${\textsc{v}}$        & '' &  $22.374$  & WA\,1   & $23.745 \pm 0.001$          &  $59 \pm 1$     &  $23.749^{+0.001}_{-0.008}$          &  $61^{+4}_{-2}$ \\
	O${\textsc{v}}$        & '' &  $22.374$  & WA\,3   &  $23.557 \pm 0.004$         &  $17 \pm 1$     & $ $                          &  $ $ \\
	O${\textsc{iv}}$       & $2p^1     \rightarrow 1s^1 2s^2 2p^2$  (K$\alpha$)  &  $22.660$  & UFO\,1  & $21.993 \pm 0.001$          &  $8 \pm 1$      &  $ $                         &  $ $  \\
	O${\textsc{iv}}$       & '' &  $22.660$  & WA\,1   & $24.049 \pm 0.001$          &  $27^{+1}_{-3}$ &  $ $                         &  $ $  \\
	O${\textsc{iv}}$       & '' &  $22.660$  & WA\,3   & $23.859 \pm 0.004$          &  $10 \pm 1$      &  $ $                         &  $ $  \\
	O${\textsc{iv}}$       & '' &  $22.740$  & UFO\,1  & $22.071 \pm 0.001$          &  $17 \pm 1$     &  $21.97 \pm 0.01$          &  $12 \pm 7$  \\
	O${\textsc{iv}}$       & '' &  $22.740$  & WA\,1   & $24.134 \pm 0.001$          &  $47 \pm 1$     &  $24.137^{+0.001}_{-0.007}$          &  $51^{+5}_{-2}$ \\
	O${\textsc{iv}}$       & '' &  $22.740$  & WA\,3   & $23.943 \pm 0.004$          &  $22 \pm 1$     &  $ $                         &  $ $ \\
	O${\textsc{iv}}$       & '' &  $22.770$  & UFO\,1  & $22.100 \pm 0.001$          &  $15 \pm 1$     &  $22.00 \pm 0.01$          &  $10 \pm 7$  \\
	O${\textsc{iv}}$       & '' &  $22.770$  & WA\,1   & $24.166 \pm 0.001$          &  $45 \pm 1$     &  $24.169^{+0.001}_{-0.008}$  &  $48^{+4}_{-1}$  \\
	O${\textsc{iv}}$       & '' &  $22.770$  & WA\,3   & $23.974 \pm 0.004$          &  $18^{+5}_{-1}$     &  $ $                         &  $ $ \\
	O${\textsc{iii}}$      & $2p^2     \rightarrow 1s^1 2s^2 2p^3$  (K$\alpha$)  &  $23.030$  & WA\,3   & $24.248 \pm 0.004$     &  $7 \pm 1$      &  $ $                         &  $ $  \\
	O${\textsc{iii}}$      & '' &  $23.070$  & UFO\,1  & $22.391 \pm 0.001$          &  $7^{+3}_{-4}$      &  $ $                         &  $ $  \\
	O${\textsc{iii}}$      & '' &  $23.070$  & WA\,1   & $ $                         &  $ $            & $24.488^{+0.001}_{-0.008}$           & $27 \pm 7$ \\
	O${\textsc{iii}}$      & '' &  $23.070$  & WA\,3   & $24.290 \pm 0.004$          &  $8 \pm 1$      &  $ $                         &  $ $  \\
	O${\textsc{iii}}$      & '' &  $23.090$  & UFO\,1  & $22.411 \pm 0.001$          &  $8 \pm 4$      &  $ $                         &  $ $  \\
	O${\textsc{iii}}$      & '' &  $23.090$  & WA\,1   & $ $                         &  $ $            & $24.509^{+0.001}_{-0.008}$           & $23 \pm 7$ \\
	O${\textsc{iii}}$      & '' &  $23.090$  & WA\,3   & $24.311 \pm 0.004$  &  $14 \pm 1$     &  $ $                         &  $ $  \\
	O${\textsc{ii}}$       & $2p^3     \rightarrow 1s^1 2s^2 2p^4$  (K$\alpha$)   &  $23.250$  & WA\,3   & $24.480 \pm 0.004$          &  $9 \pm 1$      & $24.332^{+0.004}_{-0.003}$           & $10 \pm 1$\\
	O${\textsc{ii}}$       & $2s^1 2p^4 \rightarrow 1s^1 2s^1 2p^5$ (K$\alpha$)   &  $23.292$  & WA\,3   & $24.524 \pm 0.004$          &  $21 \pm 1$     & $24.376 \pm 0.003$           & $23^{+6}_{-2}$ \\
	O${\textsc{ii}}$       & $2p^3     \rightarrow 1s^1 2s^2 2p^4$  (K$\alpha$)   &  $23.345$  & WA\,3   & $24.580 \pm 0.004$          &  $23 \pm 1$     & $24.431^{+0.004}_{-0.003}$           & $26^{+5}_{-3}$ \\
	N${\textsc{vii}}$      & $1s^1 \rightarrow 2p^1$ (K$\alpha$) &  $24.781$  & WA\,2   & $26.308 \pm 0.001$          &  $12 \pm 1$     &  $ $  &  $ $ \\
	N${\textsc{vii}}$      & '' &  $24.781$  & WA\,4   &  $ $                        &  $ $            & $26.56 \pm 0.05$           &  $7 \pm 5$ \\
	N${\textsc{vi}}$       & $1s^2 \rightarrow 1s^1 3p^1$ (K$\beta$) &  $24.898$  & WA\,1   & $26.424 \pm 0.001$          &  $26 \pm 1$     &  $ $                         &  $ $  \\
	C${\textsc{vi}}$       & $1s^1 \rightarrow 5p^1$ &  $26.357$  & UFO\,3  & $25.823 \pm 0.001 $         &  $8^{+1}_{-6}$      &  $ $                         &  $ $  \\
	C${\textsc{vi}}$       & '' &  $26.357$  & UFO\,4  & $25.823 \pm 0.001$          &  $7 \pm 1$      &  $ $                         &  $ $  \\
	C${\textsc{vi}}$       & $1s^1 \rightarrow 4p^1$ &  $26.990$  & WA\,2   & $28.653 \pm 0.001$          &  $4 \pm 1$      &  $ $                         &  $ $  \\
	C${\textsc{vi}}$       & $1s^1 \rightarrow 3p^1$ (K$\beta$) &  $28.466$  & UFO\,3  & $27.889 \pm 0.001$          &  $16^{+1}_{-8}$     &  $27.92 \pm 0.03$          &  $8 \pm 5$  \\
	C${\textsc{vi}}$       & '' &  $28.466$  & UFO\,4  & $27.889 \pm 0.001$          &  $16 \pm 1$     &  $ $                         &  $ $         \\
	C${\textsc{vi}}$       & '' &  $28.466$  & WA\,1   & $30.211 \pm 0.001$          &  $34^{+2}_{-1}$     &  $ $                         &  $ $      \\
	N${\textsc{vi}}$       & $1s^2 \rightarrow 1s^1 2p^1$ (K$\alpha$) &  $28.787$  & WA\,4   &  $ $                        &  $ $            & $30.80 \pm 0.02$           &  $26 \pm 12$ \\ 
	N${\textsc{vi}}$       & '' &  $28.787$  & UFO\,3  & $28.203 \pm 0.001$          &  $24 \pm 2$     &  $ $                         &  $ $  \\
	N${\textsc{vi}}$       & '' &  $28.787$  & UFO\,4  & $28.204 \pm 0.001$          &  $24^{+1}_{-2}$     &  $ $                         &  $ $  \\
	N${\textsc{vi}}$       & '' &  $28.787$  & WA\,1   & $30.551 \pm 0.001$          &  $57^{+2}_{-1}$     &  $30.556^{+0.001}_{-0.010}$  &  $58^{+8}_{-12}$   \\
	N${\textsc{v}}$        & $2s^1     \rightarrow 1s^1 2s^1 2p^1$ &  $29.458$  & UFO\,1  & $28.591^{+0.002}_{-0.001}$          &  $6^{+1}_{-3}$      &  $ $                         &  $ $  \\
	N${\textsc{v}}$        & '' &  $29.458$  & WA\,1   & $31.263 \pm 0.001$          &  $43 \pm 1$     &  $31.268^{+0.001}_{-0.010}$          &  $45^{+1}_{-4}$   \\
	N${\textsc{iv}}$       & $2s^2 \rightarrow 1s^1 2s^2 2p^1$      (K$\alpha$) &  $29.928$  & UFO\,1  & $29.048 \pm 0.001$          &  $10 \pm 1$     &  $28.91 \pm 0.01$          &  $9 \pm 5$  \\
	N${\textsc{iv}}$       & '' &  $29.928$  & WA\,1   & $31.762 \pm 0.002$          &  $41^{+1}_{-3}$     &  $31.767^{+0.001}_{-0.010}$          &  $42^{+6}_{-1}$ \\
	N${\textsc{iii}}$      & $2s^1 2p^2 \rightarrow 1s^1 2s^1 2p^3$  (K$\alpha$) &  $30.483$  & WA\,3   & $32.095 \pm 0.005$          &  $6 \pm 1$      &  $ $                         &  $ $ \\
	N${\textsc{iii}}$      & '' &  $30.485$  & WA\,3   & $32.098 \pm 0.006$          &  $9 \pm 1$      &  $ $                         &  $ $ \\
	N${\textsc{iii}}$      & $2p^1 \rightarrow 1s^1 2s^2 2p^2$       (K$\alpha$) &  $30.501$  & WA\,3   & $32.114 \pm 0.005$          &  $8 \pm 1$      &  $ $                         &  $ $ \\
	N${\textsc{ii}}$       & $2s^1 2p^3 \rightarrow 1s^1 2s^1 2p^4$  (K$\alpha$) &  $30.836$  & WA\,3   & $32.467 \pm 0.006$          &  $7 \pm 1$      &  $ $                         &  $ $ \\
	N${\textsc{ii}}$       & '' &  $30.879$  & WA\,3   & $32.512 \pm 0.005$          &  $8 \pm 1$      &  $ $                         &  $ $ \\
	N${\textsc{ii}}$       & $2p^2      \rightarrow 1s^1 2s^2 2p^3$  (K$\alpha$) &  $30.924$  & WA\,3   & $32.560 \pm 0.006$          &  $9 \pm 1$      &  $ $                         &  $ $ \\
	C${\textsc{v}}$        & $1s^2      \rightarrow 1s^1 5p^1$                   &  $32.754$  & WA\,1   & $34.761 \pm 0.001$          &  $35 \pm 1$     &  $34.767^{+0.001}_{-0.011}$          &  $36^{+2}_{-5}$  \\
	C${\textsc{vi}}$       & $1s^1 \rightarrow 2p^1$ (K$\alpha$) &  $33.736$  & UFO\,3  & $33.053 \pm 0.001$          &  $32^{+2}_{-24}$     &  $33.09 \pm 0.03$          &  $16 \pm 11$ \\
	C${\textsc{vi}}$       & '' &  $33.736$  & UFO\,4  & $33.053 \pm 0.001$          &  $34^{+}_{-2}$     &  $ $                         &  $ $ \\
	C${\textsc{v}}$        & $1s^2 \rightarrow 1s^1 3p^1$ (K$\beta$) &  $34.973$  & UFO\,1  & $33.944^{+0.002}_{-0.001}$          &  $12^{+1}_{-3}$     &  $ $                         &  $ $  \\
	C${\textsc{v}}$        & '' &  $34.973$  & UFO\,3  & $34.264 \pm 0.002$          &  $18 \pm 2$     &  $ $                         &  $ $  \\
	C${\textsc{v}}$        & '' &  $34.973$  & UFO\,4  & $34.264^{+0.002}_{-0.001}$          &  $17 \pm 2$     &  $ $                         &  $ $  \\
	\enddata
	\tablecomments{All the atomic transitions data are gathered from the {\small{AtomDB} database v.3.0.9} \citep{2001ApJ...556L..91S}, the {\small{NIST} database v.5.5.6} \citep{nist-database}, and the {\small{XSTAR} database v.2.39} \citep{2001ApJS..133..221K}. Every absorption line included in this table is depicted in Fig.\,\ref{fig:model}.} 
		\tablenotetext{a}{$\lambda$ refers to the theoretical wavelength of the absorption line in the source's restframe.}
		\tablenotetext{b}{The measured $\lambda$ values refer to the observer's restframe.}
\end{deluxetable}

\end{document}